\documentclass{JAC2000}
\usepackage{graphicx}
\newcommand{\ud}{\mathrm{d}}
\begin{document}
\title{MULTISCALE REPRESENTATIONS FOR SOLUTIONS OF VLASOV-MAXWELL
EQUATIONS FOR INTENSE BEAM PROPAGATION}
\author{A. Fedorova,  M. Zeitlin, 
IPME, RAS, V.O. Bolshoj pr., 61, 199178, St.~Petersburg, Russia 
\thanks{e-mail: zeitlin@math.ipme.ru}\thanks{ http://www.ipme.ru/zeitlin.html;
http://www.ipme.nw.ru/zeitlin.html}
}

\maketitle

\begin{abstract}
We present the applications of variational--wavelet approach for computing 
multiresolution/multiscale representation for solution of some
approximations of Vlasov-Maxwell equations.
\end{abstract}

\section{INTRODUCTION}
In this paper we consider the applications of a new nu\-me\-ri\-cal\--analytical 
technique which is based on the methods of local nonlinear harmonic
analysis or wavelet analysis to the nonlinear beam/accelerator physics
problems described by some forms of Vlasov-Maxwell (Poisson) equations.
Such approach may be useful in all models in which  it is 
possible and reasonable to reduce all complicated problems related with 
statistical distributions to the problems described 
by systems of nonlinear ordinary/partial differential 
equations. 
Wavelet analysis is a relatively novel set of mathematical
methods, which gives us the possibility to work with well-localized bases in
functional spaces and gives for the general type of operators (differential,
integral, pseudodifferential) in such bases the maximum sparse forms. 
Our approach in this paper is based on the generalization
of variational-wavelet 
approach from [1]-[8],
which allows us to consider not only polynomial but rational type of 
nonlinearities [9].
The solution has the following form (related forms in part 3)
\begin{eqnarray}\label{eq:z}
u(t,x)&=&\sum_{k\in Z^n}U^k(x)V^k(t),\\
V^k(t)&=&V_N^{k,slow}(t)+\sum_{j\geq N}V^k_j(\omega^1_jt), \quad \omega^1_j\sim 2^j \nonumber\\
U^k(x)&=&U_N^{k,slow}(x)+\sum_{j\geq N}U^k_j(\omega^2_jx), \quad \omega^2_j\sim 2^j \nonumber
\end{eqnarray}
which corresponds to the full multiresolution expansion in all time/space 
scales.

Formula (\ref{eq:z}) gives us expansion into the slow part $u_N^{slow}$
and fast oscillating parts for arbitrary N. So, we may move
from coarse scales of resolution to the 
finest one for obtaining more detailed information about our dynamical process.
The first term in the RHS of formulae (1) corresponds on the global level
of function space decomposition to  resolution space and the second one
to detail space. In this way we give contribution to our full solution
from each scale of resolution or each time/space scale.
The same is correct for the contribution to power spectral density
(energy spectrum): we can take into account contributions from each
level/scale of resolution.
Starting  in part 2 from Vlasov-Maxwell equations
we consider in part 3 the generalization of our approach based on
variational formulation in the biorthogonal bases of compactly
supported wavelets. 
\begin{figure}[htb]
\centering
\includegraphics*[width=50mm]{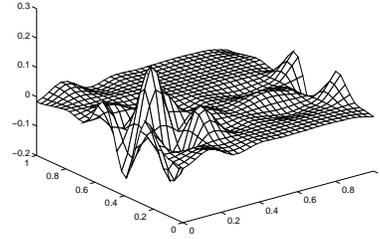}
\caption{Base wavelet.}
\end{figure}

\section{Vlasov-Maxwell Equations}

Analysis based on the non-linear Vlasov-Maxwell equations leds to more
clear understanding of the collecti\-ve effects and nonlinear beam dynamics
of high intensity beam propagation in periodic-focusing and uniform-focusing
transport systems.
We consider the following form of equations ([11] for setup and designation): 
\begin{eqnarray}
&&\Big\{\frac{\partial}{\partial s}+x'\frac{\partial}{\partial x}+
             y'\frac{\partial}{\partial y}-
\Big[k_x(s)x+\frac{\partial\psi}{\partial x}\Big]\frac{\partial}{\partial x'}-\nonumber\\
&& \Big[k_y(s)y+\frac{\partial\psi}{\partial y}\Big]\frac{\partial}{\partial y'}
  \Big\} f_b=0, \\
&&\Big(\frac{\partial^2}{\partial x^2}+\frac{\partial^2}{\partial y^2}\Big)\psi=
-\frac{2\pi K_b}{N_b}\int \ud x' \ud y' f_b.
\end{eqnarray}
The corresponding Hamiltonian for transverse sing\-le\--par\-ticle motion is given by 
\begin{eqnarray}
&&\hat H(x,y,x',y',s)=\frac{1}{2}(x'^2+y'^2) \nonumber\\
                   &&+\frac{1}{2}[k_x(s)x^2+k_y(s)y^2]+\psi(x,y,s).
\end{eqnarray}

Related Vlasov system describes longitudinal dynamics of high energy stored beam [12]: 

\begin{eqnarray}
&&\frac{\partial f}{\partial T}+v\frac{\partial f}{\partial\theta}
                               +\lambda V\frac{\partial f}{\partial v}=0,\\
&&\frac{\partial^2 V}{\partial T^2}+2\gamma\frac{\partial V}{\partial T}+
                               \omega^2 V=\frac{\partial I}{\partial T} \\
&& I(\theta;T)=\int\ud v v f(\theta,v;T). 
\end{eqnarray}

\section{Variational Approach in Biorthogonal Wavelet Bases}

Now we consider some useful generalization of our variational wavelet approach.
Because integrand of variational functionals is represented
by bilinear form (scalar product) it seems more reasonable to
consider wavelet constructions [13] which take into account all advantages of
this structure. The action functional for loops in the phase space is 
\begin{equation}
F(\gamma)=\displaystyle\int_\gamma pdq-\int_0^1H(t,\gamma(t))dt
\end{equation}
The critical points of $F$ are those loops $\gamma$, which solve
the Hamiltonian equations associated with the Hamiltonian $H$
and hence are periodic orbits.
Let us consider the loop space $\Omega=C^\infty(S^1, R^{2n})$,
where $S^1=R/{\bf Z}$, of smooth loops in $R^{2n}$.
Let us define a function $\Phi: \Omega\to R $ by setting
\begin{equation}
\Phi(x)=\displaystyle\int_0^1\frac{1}{2}<-J\dot x, x>dt-
\int_0^1 H(x(t))dt, \quad x\in\Omega
\end{equation}
Computing the derivative at $x\in\Omega$ in the direction of $y\in\Omega$,
we find
\begin{eqnarray}
\Phi'(x)(y)=
\displaystyle\int_0^1<-J\dot x-\bigtriangledown H(x),y>dt
\end{eqnarray}
Consequently, $\Phi'(x)(y)=0$ for all $y\in\Omega$ iff the loop $x$
is a solution of the Hamiltonian equations.
Now we need to take into account underlying bilinear structure via wavelets.
We started with two hierarchical sequences of approximations spaces [13]:
$
\dots V_{-2}\subset V_{-1}\subset V_{0}\subset V_{1}\subset V_{2}\dots,$
$\dots \widetilde{V}_{-2}\subset\widetilde{V}_{-1}\subset
\widetilde{V}_{0}\subset\widetilde{V}_{1}\subset\widetilde{V}_{2}\dots, 
$
and as usually,
$W_0$ is complement to $V_0$ in $V_1$, but now not necessarily orthogonal
complement.
New orthogonality conditions have now the following form:
$
\widetilde {W}_{0}\perp V_0,\quad  W_{0}\perp\widetilde{V}_{0},\quad
V_j\perp\widetilde{W}_j, \quad \widetilde{V}_j\perp W_j
$
translates of $\psi$ $\mathrm{span}$ $ W_0$,
translates of $\tilde\psi \quad \mathrm{span} \quad\widetilde{W}_0$.
Biorthogonality conditions are
$
<\psi_{jk},\tilde{\psi}_{j'k'}>=
\int^\infty_{-\infty}\psi_{jk}(x)\tilde\psi_{j'k'}(x)\ud x=
\delta_{kk'}\delta_{jj'},
$
 where
$\psi_{jk}(x)=2^{j/2}\psi(2^jx-k)$.
Functions $\varphi(x), \tilde\varphi(x-k)$ form  dual pair:
$
<\varphi(x-k),\tilde\varphi(x-\ell)>=\delta_{kl},\quad
 <\varphi(x-k),\tilde\psi(x-\ell)>=0\quad  \mbox{for}\quad \forall k,
\ \forall\ell.
$
Functions $\varphi, \tilde\varphi$ generate a multiresolution analysis.
$\varphi(x-k)$, $\psi(x-k)$ are synthesis functions,
$\tilde\varphi(x-\ell)$, $\tilde\psi(x-\ell)$ are analysis functions.
Synthesis functions are biorthogonal to analysis functions. Scaling spaces
are orthogonal to dual wavelet spaces.
Two multiresolutions are intertwining
$
V_j+W_j=V_{j+1}, \quad \widetilde V_j+ \widetilde W_j = \widetilde V_{j+1}
$.
These are direct sums but not orthogonal sums.
So, our representation for solution has now the form
\begin{equation}
f(t)=\sum_{j,k}\tilde b_{jk}\psi_{jk}(t),
\end{equation}
where synthesis wavelets are used to synthesize the function. But
$\tilde b_{jk}$ come from inner products with analysis wavelets.
Biorthogonality yields
\begin{equation}
\tilde b_{\ell m}=\int f(t)\tilde{\psi}_{\ell m}(t) \ud t.
\end{equation}
So, now we can introduce this more useful construction into
our variational approach. We have modification only on the level of
computing coefficients of reduced nonlinear algebraical system.
This new construction is more flexible.
Biorthogonal point of view is more stable under the action of large
class of operators while orthogonal (one scale for multiresolution)
is fragile, all computations are much more simpler and we accelerate
the rate of convergence. In all types of (Hamiltonian) calculation,
which are based on some bilinear structures (symplectic or
Poissonian structures, bilinear form of integrand in variational
integral) this framework leads to greater success.                               

So, we try to use wavelet bases with their good spatial and scale--wavenumber      
localization properties to explore the dynamics of coherent structures in      
spatially-extended, 'turbulent'/stochastic systems.
After some ansatzes and reductions we arrive from (2),(3) or (5)-(7)
to some system of nonlinear partial differential equations [10].
We consider application of our technique                                       
to Kuramoto-Sivashiinsky equation as a model with rich spatio-temporal behaviour [14]
($ 0\le x\le L$, $\quad\xi=x/L$, $\quad u(0,t)=u(L,t)$, 
$\quad u_x(0,t)=u_x(L,t)$):             
\begin{eqnarray}                                                               
u_t&=&-u_{xxx}-u_{xx}-uu_x=Au+B(u)\nonumber\\                                             
u_t&+&\frac{1}{L^4}u_{\xi\xi\xi\xi}+\frac{1}{L^2}u_{\xi\xi}+                   
    \frac{1}{L}uu_\xi=0
\end{eqnarray}
Let be
\begin{eqnarray}                                             
u(x,t)&=&\sum_{k=0}^N\sum_{\ell=0}^M a_{\ell}^k(t)\psi_\ell^k(\xi)=          
\sum a_\ell^k\psi_\ell^k,                                             
\end{eqnarray}                                                           
where $\psi_\ell^k(\xi)$,                           
 $a_\ell^k(t)$ are both wavelets.\\                        
Variational formulation                                                         
\begin{eqnarray}                                                              
&&\Bigg( \sum_{k,\ell} \Big\{ \dot{a}_\ell^k \psi_\ell^k                        
+\frac{1}{L^4} a_\ell^k \psi_\ell^{k{''''}} +                                   
\frac{1}{L^2}a_\ell^k \psi_\ell^{k{''}}\nonumber\\                            
&&+\frac{1}{L} \sum_{p,q}                                                     
a_\ell^k a_q^p \psi_\ell^k                                            
\psi_q^{p'}  \Big\} , \psi_s^r \Bigg)=0        
\end{eqnarray}
reduces (13) to ODE and algebraical one. 
\begin{eqnarray}
M_{s\ell}^{rk}\dot{a}_s^r&=&\sum_{k,\ell}L_{s\ell}^{rk}a_\ell^k
  +\sum_{k,\ell}\sum_{p,q}N_{sq\ell}^{rpk}a^p_q a_\ell^k \nonumber\\
M_{s\ell}^{rk}&=&\big( \psi_\ell^k,\psi_s^r \big)\\
L_{s\ell}^{rk}&=&\frac{1}{L^2}(\psi_s^{r'},\psi_\ell^{k'} \big) -
   \frac{1}{L^4}(\psi_s^{r''},\psi_\ell^{k''}) \nonumber\\
N^{rpk}_{sq\ell}&=&\frac{1}{L}(\psi_s^r,\psi_q^p\psi_\ell^{k'}) \nonumber
\end{eqnarray}
In particular case on $V_2 \setminus V_0$ we have:
\begin{displaymath}
\left(\begin{array}{l}
\dot{a}_0\\
\dot{a}_1\\
\dot{a}_2
\end{array} \right)
=\Bigg[L\Bigg]
\left(\begin{array}{l}
a_0\\
a_1\\
a_2
\end{array} \right)+
\end{displaymath}
\begin{displaymath}
\left(\begin{array}{l}
ca_0a_1-ca_0a_2+da_1^2-da_2^2\\
-ca_0^2-da_0a_1+\ell a_0a_2-fa_1a_2-fa_2^2\\
ca_0^2-\ell a_0a_1+da_0a_1+da_0a_2+fa_1^2+fa_1a_2
\end{array}\right)
\end{displaymath}
Then in contrast to [14] we apply to (16) methods from [1]-[9] and arrive to formula (1).
The same approach we use for the general nonlinear wave equation
\begin{eqnarray}
u_{tt}=u_{xx}-mu-f(u),
\end{eqnarray}
where
\begin{eqnarray}
f(u)=au^3+\sum_{k\ge 5} f_ku^k
\end{eqnarray}
According to [2],[10] we may consider it as infinite dimensional Hamiltonian systems with
phase space $=H_0^1\times L^2$ on $[0, L]$
and coordinates: $u, v=u_t$, then
\begin{eqnarray}
H&=&\frac{1}{2}<v,v>+\frac{1}{2}<Au,u>+\int_0^\pi g(u)\textrm{d}x \nonumber\\
A&=&\frac{\textrm{d}^2}{\textrm{d}x^2}+m,\qquad g=\int f(s)\textrm{d}s\nonumber\\
u_t&=&\frac{\partial H}{\partial v}=v\\
v_t&=&-\frac{\partial H}{\partial u}=-Au-f(u)\nonumber\\
\textrm{or} \quad \dot{u}(t)&=&J\nabla K(u(t))\nonumber
\end{eqnarray}
Then anzatzes:
\begin{eqnarray}
u(t,x)&=&U(\omega_1t,\dots,\omega_nt,x)\\
u(t,x)&=&\sum_{k\in\mathcal{Z}^n}U_k(x)\exp(ik\cdot\omega(k) t)\nonumber\\
u(t,x)&=&S(x-vt)\nonumber\\
u(t,x)&=&\sum_{k\in\mathcal{Z}^n}U_k(x)V_k(t)\nonumber
\end{eqnarray}
and methods [1]-[10] led to formulae (1).
Resulting multiresolution/multiscale representation in the high-localized bases
(Fig.1) is demonstrated on Fig.2, Fig.3.
\begin{figure}
\centering
\includegraphics*[width=60mm]{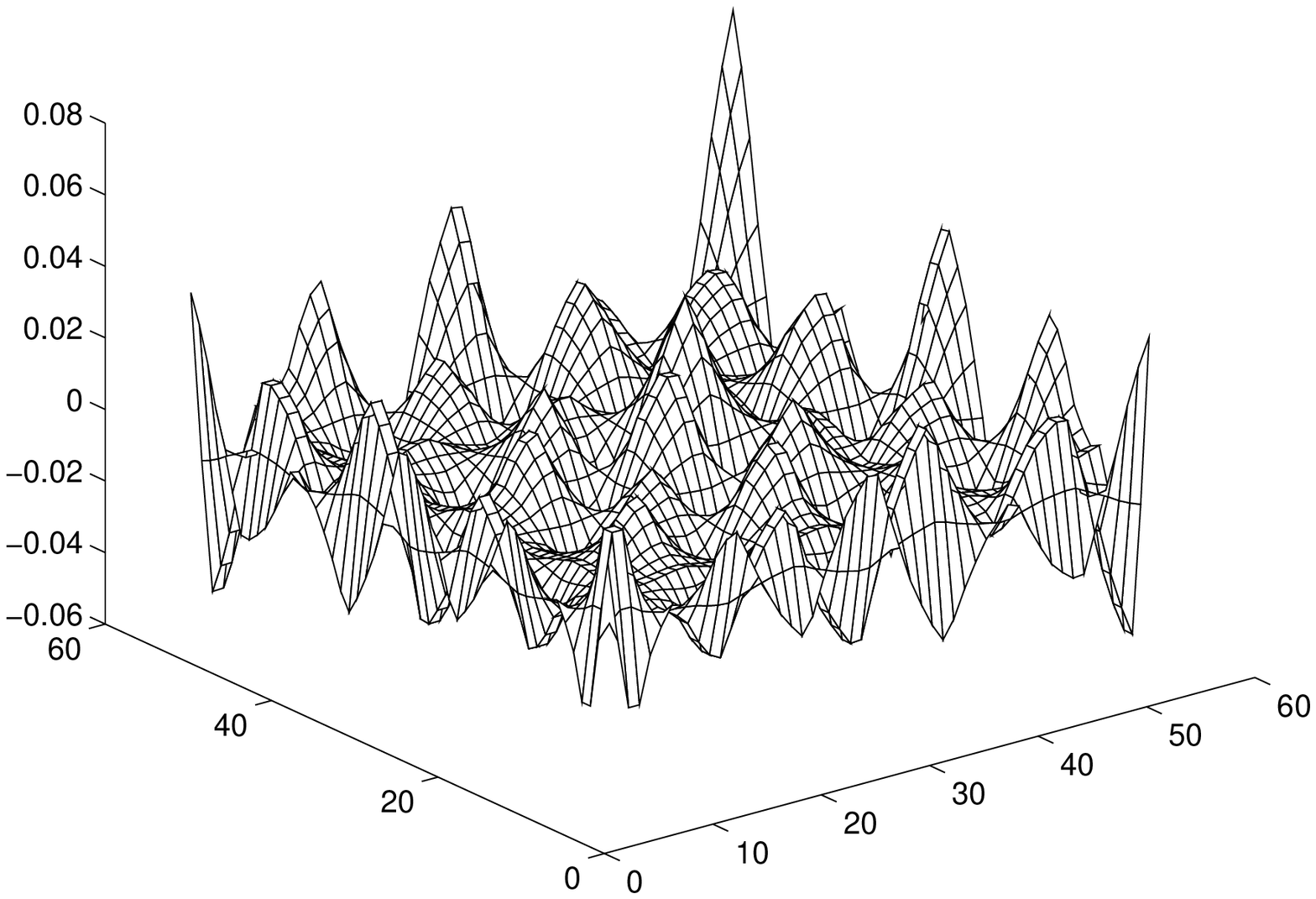}
\caption{The solution of eq.(13)}
\end{figure}
\begin{figure}
\centering
\includegraphics*[width=60mm]{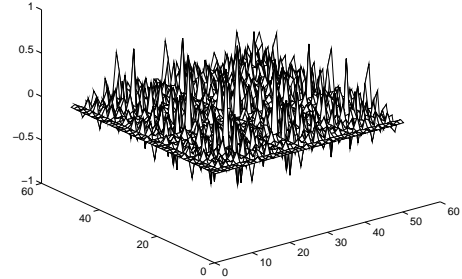}
\caption{The solution of eq.(17)}
\end{figure}
We would like to thank Professor
James B. Rosenzweig and Mrs. Melinda Laraneta for
nice hospitality, help and support during UCLA ICFA Workshop.


\begin{thebibliography}{14}



\bibitem{1}
A.N. Fedorova and M.G. Zeitlin, 
'Wavelets in Optimization and Approximations', 
 {\it Math. and Comp. in Simulation}, {\bf 46}, 527, 1998.


\bibitem{2}
A.N. Fedorova and M.G. Zeitlin,
'Wavelet Approach to Mechanical Problems. Symplectic Group,
Symplectic Topology and Symplectic Scales',
{\it New Applications of Nonlinear and Chaotic Dynamics in Mechanics}, 31, 101
(Kluwer,  1998).

\bibitem{3}
A.N. Fedorova and M.G. Zeitlin,
 'Nonlinear Dynamics of Accelerator via Wavelet
Approach', {\bf CP405}, 87 (American Institute of Physics, 1997).\\
Los Alamos preprint, phy\-sics/9710035.

\bibitem{4}
A.N. Fedorova, M.G. Zeitlin and Z.~Parsa, 
'Wavelet Approach to Accelerator Problems', parts 1-3, Proc. PAC97 
{\bf 2}, 1502, 1505, 1508 (IEEE, 1998).

\bibitem{5}
A.N. Fedorova, M.G. Zeitlin and Z.~Parsa, 
Proc. EPAC98, 930, 933 (Institute of Physics, 1998).

\bibitem{6}
A.N. Fedorova, M.G. Zeitlin and Z.~Parsa,    
Variational Approach in
    Wavelet Framework to Polynomial
    Approximations of Nonlinear Accelerator Problems.
    {\bf CP468}, 48 ( American Institute of Physics, 1999).\\
Los Alamos preprint, physics/990262

\bibitem{7}
A.N. Fedorova, M.G. Zeitlin and Z.~Parsa,  
Symmetry, Hamiltonian
 Problems and Wavelets in
    Accelerator Physics.
    {\bf CP468}, 69 (American Institute of Physics, 1999).\\
Los Alamos preprint, physics/990263

 \bibitem{8}
A.N. Fedorova and M.G. Zeitlin,  
Nonlinear Accelerator Problems
     via Wavelets, parts 1-8,
     Proc. PAC99, 
     1614, 1617, 1620, 2900, 2903,
      2906, 2909, 2912 (IEEE/APS, New York, 1999).\\
Los Alamos preprints: 
physics/9904039,  physics/9904040, physics/\-9904041, physics\-/\-9904042,
physics/9904043, phy\-sics/9904045, physics/9904046, physics/9904047.

\bibitem{9}
A.N. Fedorova and M.G. Zeitlin,
Los Alamos preprint: physics/0003095

\bibitem{10}
A.N. Fedorova and M.G. Zeitlin,in press

\bibitem{11}
R. Davidson, H. Qin, P. Channel,
PRSTAB, 2, 074401, 1999

\bibitem{12}
S. Tzenov, P. Colestock,
Fermilab-Pub-98/258

\bibitem{13}
A. Cohen, I. Daubechies and J.C. Feauveau,  {\it Comm. Pure. Appl. Math.},
 {\bf XLV}, 485 (1992).

\bibitem{14}
Ph. Holmes e.a., Physica D86, 396, 1995

\end{thebibliography}
 \end{document}